\documentclass[aps,twocolumn,amsmath,letterpaper,floatfix]{revtex4}
\usepackage{amssymb}
\usepackage{graphicx}
\usepackage{array}
\usepackage{hhline}
\usepackage{longtable}

\renewcommand{\thetable}{\Roman{table}} \thetable

\begin{document}

\title{$d=3$ Anisotropic and $d=2$ $tJ$ Models:

Phase Diagrams, Thermodynamic Properties, and Chemical Potential
Shift}

\author{Michael Hinczewski$^{1,2}$ and A. Nihat Berker$^{1-4}$}

\affiliation{$^1$Feza G\"ursey Research Institute, T\"UBITAK -
Bosphorus University, \c{C}engelk\"oy 34680, Istanbul, Turkey,}

\affiliation{$^2$Department of Physics, Massachusetts Institute of
Technology, Cambridge, Massachusetts 02139, U.S.A.,}

\affiliation{$^3$Department of Physics, Ko\c{c} University, Sar\i
yer 34450, Istanbul, Turkey,}

\affiliation{$^4$Department of Physics, Istanbul Technical
University, Maslak 34469, Istanbul, Turkey}

\begin{abstract}
  The anisotropic $d=3$ \:$tJ$ model is studied by
  renormalization-group theory, yielding the evolution of the system
  as interplane coupling is varied from the isotropic
  three-dimensional to quasi-two-dimensional regimes.
  Finite-temperature phase diagrams, chemical potential shifts, and
  in-plane and interplane kinetic energies and antiferromagnetic
  correlations are calculated for the entire range of electron
  densities. We find that the novel $\tau$ phase, seen in earlier
  studies of the isotropic $d=3$ \:\:$tJ$ model, persists even for
  strong anisotropy. While the $\tau$ phase appears at low
  temperatures at $30-35\%$ hole doping away from $\langle
  n_i\rangle=1$, at smaller hole dopings we see a complex lamellar
  structure of antiferromagnetic and disordered regions, with a
  suppressed chemical potential shift, a possible marker of
  incommensurate ordering in the form of microscopic stripes.  An
  investigation of the renormalization-group flows for the isotropic
  two-dimensional $tJ$ model also shows a clear pre-signature of the
  $\tau$ phase, which in fact appears with finite transition
  temperatures upon addition of the smallest interplane coupling.

\end{abstract}
\maketitle
\def\s{\rule{0in}{0.28in}}

\section{Introduction}
\setlength{\LTcapwidth}{\columnwidth}

The anisotropic nature of high-$T_c$ materials, where groups of one or
more CuO$_2$ planes are weakly coupled through block layers that act
as charge reservoirs, has led to intense theoretical focus on
two-dimensional models of electron conduction.\cite{ADagotto} However,
a full understanding of the cuprates will require taking into account
physics along the third dimension.  Crucial aspects of the phase
diagram, like the finite value of the N\'eel temperature, depend on
interplanar coupling~\cite{AImada}, and going beyond two dimensions is
also necessary to explain the behavior of $T_c$ as the number of
CuO$_2$ layers per unit cell is increased~\cite{AChakravarty}.
Moreover, given the recent debate over the adequacy of the
two-dimensional $tJ$ model as a description of high-$T_c$
superconductivity~\cite{PryadkoKivelsonZachar,KoretsuneOgata,PutikkaLuchini,
  Su}, a resolution of the issue might be found by turning to highly
anisotropic three-dimensional models~\cite{Su}.

As a simplified description of strongly correlated electrons in an
anisotropic system, we look at the $tJ$ model on a cubic lattice
with uniform interaction strengths in the $xy$ planes, and a weaker
interaction in the $z$ direction.  To obtain a finite-temperature
phase diagram for the entire range of electron densities, we extend
to anisotropic systems the renormalization-group approach that has
been applied successfully in earlier studies of both $tJ$ and
Hubbard models as isotropic $d=3$ systems.\cite{AFalicovBerkerT,
AFalicovBerker,AHinczBerker1,AHinczBerker2}  For the $d=3$ isotropic
$tJ$ model, this approach has yielded an interesting phase diagram
with antiferromagnetism near $\langle n_i\rangle=1$ and a new
low-temperature ``$\tau$'' phase for 33-37\% hole doping.  Within
this $\tau$ phase, the magnitude of the electron hopping strength in
the Hamiltonian tends to infinity as the system is repeatedly
rescaled.\cite{AFalicovBerker} The calculated superfluid
weight shows a marked peak in the $\tau$ phase, and both the
temperature profile of the superfluid weight and the density of free
carriers with hole doping is reminiscent of experimental results in
cuprates.\cite{AHinczBerker2} Given these apparent links with
cuprate physics, the next logical step is to ask whether the $\tau$
phase is present in the strongly anisotropic regime, which is the
one directly relevant to experiments.

The extension of the position-space renormalization-group
method to spatial anisotropy has recently been demonstrated with
$d=3$ Ising, XY magnetic and percolation
systems.\cite{AErbasTuncerYucesoyBerker} We apply a similar
anisotropic generalization to the electronic conduction model and find
the evolution of the phase diagram from the isotropic $d=3$ to the
quasi $d=2$ cases.  While transition temperatures become lower, the
$\tau$ phase does continue to exist even for very weak interplanar
coupling. The density range of the $\tau$ phase remains stable as
anisotropy is increased, while for 5-30\% hole doping an intricate
structure of antiferromagnetic and disordered phases develops, a
possible indicator of underlying incommensurate order, manifested
through the formation of microscopic stripes.  Consistent with this
interpretation, our system in this density range shows a
characteristic ``pinning'' of the chemical potential with hole doping.

Lastly, we turn from the $d=3$ anisotropic case to the $d=2$ $tJ$
model, where earlier studies ~\cite{AFalicovBerkerT,AFalicovBerker}
have found no $\tau$ phase (but have elucidated the
occurrence/non-occurrence of phase separation).  Nevertheless, by
looking at the low-temperature behavior of the renormalization-group
flows, we find a compelling pre-signature of the $\tau$ phase even
in $d=2$, at exactly the density range where the $\tau$ phase
appears when the slightest interplanar coupling is added to the
system.

\section{Anisotropic $tJ$ Hamiltonian}

We consider the $tJ$ Hamiltonian on a cubic lattice with different
interaction strenghts for nearest neighbors lying in the $xy$ plane
or along the $z$ direction (respectively denoted by $\langle i j
\rangle_{xy}$ and $\langle i j \rangle_z$):
\begin{equation}
\label{anis_eq:1}\begin{split} H &= P \left[ \tilde{t}_{xy} \sum_{\langle
ij \rangle_{xy},\sigma} \left(c^\dagger_{i\sigma}c_{j\sigma} +
c^\dagger_{j\sigma}c_{i\sigma}\right)\right.\\
&\qquad + \tilde{t}_{z} \sum_{\langle ij \rangle_{z},\sigma}
\left(c^\dagger_{i\sigma}c_{j\sigma} +
c^\dagger_{j\sigma}c_{i\sigma}\right)\\
&\qquad + \tilde{J}_{xy} \sum_{\langle ij \rangle_{xy}}
\mathbf{S}_i\cdot\mathbf{S}_j
+\tilde{J}_{z} \sum_{\langle ij \rangle_{z}} \mathbf{S}_i\cdot\mathbf{S}_j\\
&\qquad \left.- \tilde{V}_{xy} \sum_{\langle ij \rangle_{xy}} n_i
n_j - \tilde{V}_{z} \sum_{\langle ij \rangle_{z}} n_i n_j -
\tilde{\mu} \sum_i n_i \right] P\,.\end{split}
\end{equation}
Here $c^\dagger_{i\sigma}$ and $c_{i\sigma}$ are creation and
annihilation operators, obeying anticommutation rules, for an
electron with spin $\sigma =\: \uparrow$ or $\downarrow$ at lattice
site $i$, $n_{i\sigma} = c^\dagger_{i\sigma}c_{i\sigma}$, $n_i =
n_{i\uparrow} + n_{i\downarrow}$ are the number operators, and
$\mathbf{S}_i = \frac{1}{2}\sum_{\sigma\sigma^\prime} c^\dagger_{i\sigma}
\mathbf{s}_{\sigma\sigma^\prime} c_{i\sigma^\prime}$ is the
single-site spin operator, with $\mathbf{s}$ the vector of Pauli
spin matrices.  The entire Hamiltonian is sandwiched between
projection operators $P = \prod_{i}
(1-n_{i\downarrow}n_{i\uparrow})$, which project out states with
doubly-occupied sites.  The standard, isotropic $tJ$ Hamiltonian
obtains when $\tilde{t}_{xy} = \tilde{t}_{z}$, $\tilde{J}_{xy} =
\tilde{J}_{z}$, $\tilde{V}_{xy} = \tilde{V}_{z}$, and
$\tilde{V}_{xy} /\tilde{J}_{xy} = \tilde{V}_z/\tilde{J}_z = 1/4$.

For simplicity, we rewrite Eq.~\eqref{anis_eq:1} using dimensionless
interaction constants, and rearrange the $\tilde{\mu}$ chemical
potential term to group the Hamiltonian into summations over the
$xy$ and $z$ bonds:
\begin{equation}
\label{anis_eq:2} \begin{split} -\beta H = & \sum_{\langle ij
\rangle_{xy}} P \biggl[ -t_{xy} \sum_{\sigma}
\left(c^\dagger_{i\sigma}c_{j\sigma} +
c^\dagger_{j\sigma}c_{i\sigma}\right)\\
&\qquad- J_{xy}
\mathbf{S}_i\cdot\mathbf{S}_j + V_{xy} n_i n_j + \mu ( n_i
+n_j)\biggr] P\\
+ &\sum_{\langle ij \rangle_{z}} P \biggl[ -t_{z} \sum_{\sigma}
\left(c^\dagger_{i\sigma}c_{j\sigma} +
c^\dagger_{j\sigma}c_{i\sigma}\right)\\
&\qquad - J_{z}
\mathbf{S}_i\cdot\mathbf{S}_j + V_{z} n_i n_j + \mu ( n_i
+n_j)\biggr] P \\
\equiv &\sum_{\langle ij \rangle_{xy}} \{-\beta H_{xy}
(i,j)\}+\sum_{\langle ij \rangle_{z}} \{-\beta H_{z} (i,j)\}\,.
\end{split}
\end{equation}
Here $\beta = 1/k_B T$, so that the interaction constants are
related by $t_{xy} = \beta \tilde{t}_{xy}$,  $t_{z} = \beta
\tilde{t}_{z}$, $J_{xy} = \beta \tilde{J}_{xy}$, $J_{z} = \beta
\tilde{J}_{z}$, $V_{xy}= \beta \tilde{V}_{xy}$, $V_{z}= \beta
\tilde{V}_{z}$, and $\mu = \beta\tilde{\mu}/6$.

\section{Renormalization-Group Theory}

\subsection[Isotropic Transformation and Anisotropic\\ Expectations]{Isotropic Transformation and Anisotropic Expectations}

Since the isotropic model is a special case of Eq.~\eqref{anis_eq:1}, let
us briefly outline the main steps in effecting the renormalization
equations of earlier, isotropic
studies~\cite{AFalicovBerker,AFalicovBerkerT,AHinczBerker2}.  We begin
by setting up a decimation transformation for a one-dimensional $tJ$
chain, finding a thermodynamically equivalent Hamiltonian by tracing
over the degrees of freedom at every other lattice site.  With the
vector $\mathbf{K}$ whose elements are the interaction constants in
the Hamiltonian, the decimation can be expressed as a mapping of the
original $d=1$ system onto a new system with interaction constants
\begin{equation}\label{anis_eq:3}
\mathbf{K}^\prime = \mathbf{R}(\mathbf{K})\,.
\end{equation}
The Migdal-Kadanoff~\cite{AMigdal,AKadanoff} procedure has been
remarkably successful, for systems both classical and quantum, in
extending this transformation to $d> 1$ (for an overview, see~
\cite{AHinczBerker1}). In this procedure, a subset of the
nearest-neighbor interactions in the lattice are ignored, leaving
behind a new $d$-dimensional hypercubic lattice where each point is
connected to its neighbor by two consecutive nearest-neighbor
segments of the original lattice. The decimation described above is
applied to the middle site between the two consecutive segments,
giving the renormalized nearest-neighbor couplings for the points
forming the new lattice.  We compensate for the interactions that
are ignored in the original lattice by multiplying the interactions
after the decimation by $b^{d-1}$, where $b=2$ is the length
rescaling factor. Thus for $d>1$ the renormalization-group
transformation of Eq.~\eqref{anis_eq:3} generalizes to
\begin{equation} \label{anis_eq:4}
\mathbf{K}^\prime=b^{d-1} \mathbf{R}(\mathbf{K}),
\end{equation}
which, through flows in a four-dimensional Hamiltonian space (for
the Hubbard model, 10-dimensional Hamiltonian
space~\cite{AHinczBerker1}), yields a rich array of physical
phenomena.

With the anisotropic $tJ$ Hamiltonian on a cubic lattice
(Eq.~\eqref{anis_eq:1}), there are two intercoupled sets of interaction
constants, $\mathbf{K}_{xy}$ and $\mathbf{K}_{z}$, and further
development of the transformation is needed.  However, there are
three particular instances where the transformation in
Eq.~\eqref{anis_eq:4} is directly applicable. When $\mathbf{K}_{xy} =
\mathbf{K}_{z}$, we have the $d=3$ isotropic case, so the
appropriate renormalization-group equations are
\begin{equation}\label{anis_eq:5}
\mathbf{K}^\prime_{xy} = 4\, \mathbf{R}(\mathbf{K}_{xy})\,, \qquad
\mathbf{K}^\prime_{z} = 4\, \mathbf{R}(\mathbf{K}_{z})\,.
\end{equation}
When $\mathbf{K}_{xy} \ne 0$ and $\mathbf{K}_{z} = 0$, we have a
system of decoupled isotropic $d=2$ planes, and the transformation
is given by
\begin{equation}\label{anis_eq:6}
\mathbf{K}^\prime_{xy} = 2\, \mathbf{R}(\mathbf{K}_{xy})\,, \qquad
\mathbf{K}^\prime_{z} = 0\,.
\end{equation}
Similarly, when $\mathbf{K}_{xy} = 0$ and $\mathbf{K}_{z} \ne 0$, we
have decoupled $d=1$ chains, and
\begin{equation}\label{anis_eq:7}
\mathbf{K}^\prime_{xy} = 0\,, \qquad \mathbf{K}^\prime_{z} =
\mathbf{R}(\mathbf{K}_{z})\,.
\end{equation}
The renormalization-group transformation for the anisotropic model
described in the following sections recovers the correct results,
Eqs.\eqref{anis_eq:5}-\eqref{anis_eq:7}, for these three cases. \vspace{0em}

\begin{figure}[!t]
\centering \includegraphics*[scale=0.5]{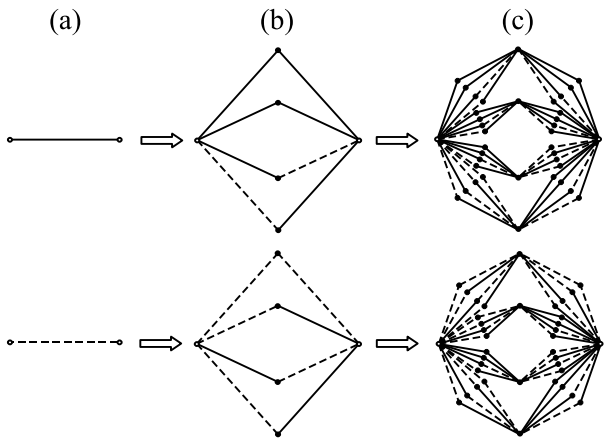}
\caption[Construction of the hierarchical model]{Construction of the
hierarchical model in Section II.B.  Solid lines correspond
to $xy$ bonds, while dashed lines correspond to $z$ bonds.}\label{anis_fig:1}
\end{figure}

\subsection{Hierarchical Lattice Model for Anisotropy}

A one-to-one correspondance exists between Migdal-Kadanoff and other
approximate renormalization-group transformations on the one hand,
and exact renormalization-group transformations of corresponding
hierarchical lattices on the other hand, through the sharing of
identical recursion
relations.\cite{ABerkerOstlund,AGriffithsKaufman,Kaufman2} The
correspondance guarantees the fulfilment of general physical
preconditions on the results of approximate renormalization-group
transformations, since the latter are thus ``physically
realizable''.\cite{ABerkerOstlund} This correspondance has recently
been exploited to develop renormalization-group transformations for
spatially anisotropic Ising, XY magnetic and percolation
systems.\cite{AErbasTuncerYucesoyBerker}  Similarly, to derive an
approximate renormalization-group transformation for the anisotropic
$tJ$ Hamiltonian, consider the nonuniform hierarchical model
depicted in Fig.~\ref{anis_fig:1}. The two types of bonds in the
lattice, corresponding to $xy$ and $z$ bonds, are drawn with solid
and dashed lines respectively. The hierarchical model is constructed
by replacing each single bond of a given type with the connected
cluster of bonds shown in Fig.~\ref{anis_fig:1}(b), and repeating
this step an arbitrary number of times. Fig.~\ref{anis_fig:1}(c)
shows the next stage in the construction for the two graphs in
column (b).  The renormalization-group transformation on this
hierarchical lattice consists of decimating over the four inner
sites in each cluster, to generate a renormalized interaction
between the two outer sites, thus reversing the construction
process, going from the graphs in column (b) of
Fig.~\ref{anis_fig:1} to those in column (a). This
renormalization-group transformation has the desired feature that in
all three of the cases described above, it reproduces the various
isotropic recursion relations of
Eqs.~\eqref{anis_eq:5}-\eqref{anis_eq:7}. \vspace{1em}

\subsection[Renormalization-Group Equations for Anisotropic\\ System]{Renormalization-Group Equations for Anisotropic System}

The hierarchical lattice can be subdivided into individual clusters of
bonds shown in Fig.~\ref{anis_fig:1}(b). We label these two types of
clusters the ``$xy$ cluster'' (Fig.~\ref{anis_fig:1}(b) top) and the
``$z$ cluster'' (Fig.~\ref{anis_fig:1}(b) bottom).  The sum over $\langle i j
\rangle_\text{$xy$ clus}$ denotes a sum over the outer sites of all
the $xy$ clusters, and analogously $\langle i j \rangle_\text{$z$
clus}$ denotes a sum over the outer sites of all $z$ clusters.  For a
given cluster with outer sites $ij$, the associated inner sites are
labeled $k^{(ij)}_1,\ldots,k^{(ij)}_4$. Then the $tJ$ Hamiltonian on
the anisotropic lattice has the form

\begin{widetext}
\begin{equation}\label{anis_eq:8}
\begin{split} -\beta H &= \sum_{\langle ij \rangle_\text{$xy$ clus}}
\Bigl[-\beta H_{xy} (i,k^{(ij)}_1)-\beta H_{xy} (k^{(ij)}_1,j)-\beta
H_{xy} (i,k^{(ij)}_2)-\beta H_{xy}
(k^{(ij)}_2,j)\\
&\qquad-\beta H_{xy} (i,k^{(ij)}_3)-\beta H_{z} (k^{(ij)}_3,j)-\beta
H_{z} (i,k^{(ij)}_4)-\beta H_{xy}
(k^{(ij)}_4,j)\Bigr]\\
&+ \sum_{\langle ij \rangle_\text{$z$ clus}} \Bigl[-\beta H_{z}
(i,k^{(ij)}_1)-\beta H_{z} (k^{(ij)}_1,j)-\beta H_{z}
(i,k^{(ij)}_2)-\beta H_{xy}
(k^{(ij)}_2,j)\\
&\qquad-\beta H_{xy} (i,k^{(ij)}_3)-\beta H_{z} (k^{(ij)}_3,j)-\beta
H_{z} (i,k^{(ij)}_4)-\beta H_{xy} (k^{(ij)}_4,j)\Bigr]\,.
\end{split}
\end{equation}
\end{widetext}

The renormalization-group transformation consists of finding a
thermodynamically equivalent Hamiltonian $-\beta^\prime H^\prime$
that involves only the outer sites of each cluster. Since we are
dealing with a quantum system, the non-commutation of the operators
in the Hamiltonian means that this decimation, tracing over the
degrees of freedom at the $k$ sites, can only be carried out
approximately~\cite{ASuzTak,ATakSuz}:

\begin{widetext}
\begin{equation}\label{anis_eq:9}
\begin{split}
\text{Tr}_{\text{$k$ sites}}\,e^{-\beta H}&\simeq \prod_{\langle ij
\rangle_\text{$xy$ clus}} \Bigl[\text{Tr}_{k^{(ij)}_1}\,e^{-\beta
H_{xy} (i,k^{(ij)}_1)-\beta H_{xy}
(k^{(ij)}_1,j)}\,\text{Tr}_{k^{(ij)}_2}\,e^{-\beta H_{xy}
(i,k^{(ij)}_2)-\beta H_{xy}
(k^{(ij)}_2,j)}\\
&\qquad\qquad\text{Tr}_{k^{(ij)}_3}\,e^{-\beta H_{xy}
(i,k^{(ij)}_3)-\beta H_{z}
(k^{(ij)}_3,j)}\,\text{Tr}_{k^{(ij)}_4}\,e^{-\beta H_{z}
(i,k^{(ij)}_4)-\beta H_{xy}
(k^{(ij)}_4,j)}\Bigr]\\
&\quad \cdot \prod_{\langle ij \rangle_\text{$z$ clus}}
\Bigl[\text{Tr}_{k^{(ij)}_1}\,e^{-\beta H_{z} (i,k^{(ij)}_1)-\beta
H_{z} (k^{(ij)}_1,j)}\,\text{Tr}_{k^{(ij)}_2}\,e^{-\beta H_{z}
(i,k^{(ij)}_2)-\beta H_{xy}
(k^{(ij)}_2,j)}\\
&\qquad\qquad\text{Tr}_{k^{(ij)}_3}\,e^{-\beta H_{xy}
(i,k^{(ij)}_3)-\beta H_{z}
(k^{(ij)}_3,j)}\,\text{Tr}_{k^{(ij)}_4}\,e^{-\beta H_{z}
(i,k^{(ij)}_4)-\beta
H_{xy} (k^{(ij)}_4,j)}\Bigr]\\
&= \prod_{\langle ij \rangle_\text{$xy$ clus}} \Bigl[
e^{-\beta^\prime H^\prime_{xy,xy}(i,j)} e^{-\beta^\prime
H^\prime_{xy,xy}(i,j)} e^{-\beta^\prime
H^\prime_{xy,z}(i,j)}e^{-\beta^\prime
H^\prime_{z,xy}(i,j)}\Bigr]\\
&\quad \cdot \prod_{\langle ij \rangle_\text{$z$ clus}} \Bigl[
e^{-\beta^\prime H^\prime_{z,z}(i,j)} e^{-\beta^\prime
H^\prime_{xy,z}(i,j)}e^{-\beta^\prime H^\prime_{z,xy}(i,j)} e^{-\beta^\prime H^\prime_{z,xy}(i,j)}\Bigr]\\
&\simeq e^{\sum_{\langle ij \rangle_\text{$xy$ clus}}
\left[-2\beta^\prime H^\prime_{xy,xy}(i,j) -\beta^\prime
H^\prime_{xy,z}(i,j)-\beta^\prime
H^\prime_{z,xy}(i,j)\right]+\sum_{\langle ij \rangle_\text{$z$
clus}} \left[ -\beta^\prime H^\prime_{z,z}(i,j)-\beta^\prime
H^\prime_{xy,z}(i,j)-2\beta^\prime H^\prime_{z,xy}(i,j)\right]}\\
&= e^{\sum_{\langle ij \rangle_\text{$xy$ clus}} \left[-\beta^\prime
H_{xy}^\prime(i,j)\right]+\sum_{\langle ij \rangle_\text{$z$ clus}}
\left[ -\beta^\prime H^\prime_z(i,j)\right]}\ = e^{-\beta^\prime
H^\prime}\,.
\end{split}
\end{equation}
\end{widetext}
Here $-\beta^\prime H^\prime_{A,B}(i,j)$, where $A$, $B$ can each be
either $xy$ or $z$, is
\begin{equation}\label{anis_eq:10}
e^{-\beta^\prime H^\prime_{A,B}(i,j)} = \text{Tr}_{k}\,e^{-\beta
H_{A} (i,k)-\beta H_{B} (k,j)}\,.
\end{equation}
In the two approximate steps, marked by $\simeq$ in
Eq.~\eqref{anis_eq:9}, we ignore the non-commutation of operators outside
three-site segments of the unrenormalized system. (On the other
hand, anticommutation rules are correctly accounted for within the
three-site segments, at all successive length scales in the
iterations of the renormalization-group transformation.) These two
steps involve the same approximation but in opposite directions,
which gives some mutual compensation. This approach has been shown
to successfully predict finite-temperature behavior in earlier
studies~\cite{ASuzTak,ATakSuz}.

Derivation of the renormalization-group equations involves
extracting the algebraic form of the operators $-\beta^\prime
H^\prime_{A,B}(i,j)$ from Eq.~\eqref{anis_eq:10}. Since $e^{-\beta^\prime
H^\prime_{A,B}(i,j)}$ and $e^{-\beta H_{A} (i,k)-\beta H_{B} (k,j)}$
act on the space of two-site and three-site states respectively,
Eq.~\eqref{anis_eq:10} can be rewritten in terms of matrix elements as
\begin{equation}\label{anis_eq:12}\begin{split}
&\langle u_i v_j | e^{-\beta^\prime H^\prime_{A,B}(i,j)}|
\bar{u}_i \bar{v}_j \rangle\\
&\qquad = \sum_{w_k} \langle u_i w_k v_j | e^{-\beta H_{A}
(i,k)-\beta H_{B} (k,j)} |\bar{u}_i w_k \bar{v}_j \rangle\,,
\end{split}
\end{equation}
where $u_{i},w_{k},v_{j},\bar{u}_{i},\bar{v}_{j}$ are single-site
state variables. Eq.(\ref{anis_eq:12}) is the contraction of a $27\times
27$ matrix on the right into a $9\times 9$ matrix on the left.  We
block-diagonalize the left and right sides of Eq.(\ref{anis_eq:12}) by
choosing basis states which are the eigenstates of total particle
number, total spin magnitude, total spin $z$-component, and parity.
We denote the set of 9 two-site eigenstates by $\{|\phi _{p}\rangle
\}$ and the set of 27 three-site eigenstates by $\{|\psi _{q}\rangle
\}$, and list them in Tables~\ref{anis_tab:1} and \ref{anis_tab:2}.  Eq.(\ref{anis_eq:12}) is
rewritten as
\begin{equation}\label{anis_eq:13}
\begin{split}
&\langle \phi _{p}|e^{-\beta^{\prime }H_{A,B}^{\prime }(i,j)}|\phi _{\bar{p}%
}\rangle = \\
&\sum_{\substack{u,v,\bar{u},\\ \bar{v},w}}
\sum_{\substack{q,\bar{q}}} \langle\phi _p|u_i v_j\rangle \langle
u_i w_k v_j|\psi_q\rangle\\ &\quad \cdot \langle \psi _q|e^{-\beta H_A(i,k)-\beta
H_B(k,j)}|\psi _{\bar{q}}\rangle \langle\psi_{\bar{q}}|\bar{u}_i w_k \bar{v}_j\rangle \langle
\bar{u}_i \bar{v}_j|\phi _{\bar{p}}\rangle\:.
\end{split}
\end{equation}

Eq.~\eqref{anis_eq:13} yields six independent elements for the matrix
$\langle \phi _{p}|e^{-\beta ^{\prime }H_{A,B}^{\prime
}(i,j)}|\phi_{\bar{p}}\rangle$, labeled $\gamma_p$ as follows:
\begin{equation}\label{anis_eq:14}
\begin{split}
\gamma_p &\equiv \langle \phi _{p}|e^{-\beta^{\prime
}H_{A,B}^{\prime
}(i,j)}|\phi_{p}\rangle \quad \text{for}\: p = 1,2,4,6,7,\\
\gamma_0 &\equiv \langle \phi _{2}|e^{-\beta^{\prime
}H_{A,B}^{\prime }(i,j)}|\phi_{4}\rangle\,.
\end{split}
\end{equation}
The number of $\gamma_p$ is also the number of interaction strengths
that are independently fixed in the Hamiltonian $-\beta ^{\prime
}H_{A,B}^{\prime }(i,j)$, which consequently must have a more
general form than the two-site Hamiltonians in Eq.~\eqref{anis_eq:2}. The
generalized form of the pair Hamiltonian is
\begin{equation}
\label{anis_eq:15}\begin{split} -\beta H(i,j) &= P \biggl[ -t
\sum_{\sigma} \left(c^\dagger_{i\sigma}c_{j\sigma}+
c^\dagger_{j\sigma}c_{i\sigma}\right)\\
&\qquad - J \mathbf{S}_i\cdot\mathbf{S}_j
+ V n_i n_j\\
&\qquad + \mu ( n_i +n_j) + \nu(n_i - n_j)+G\biggr] P
\end{split}
\end{equation}
The new terms here are: $G$, the additive constant that appears in
all renormalization-group calculations, does not affect the flows,
but enters the determination of expectation values; and $\nu(n_i -
n_j)$, a staggered term arising from decimation across two
consecutive bonds of different strengths.  Provisions for handling
the $\nu$ term will be described later in this section.

To calculate the $\gamma_p$, we determine the matrix elements of
$-\beta H_A(i,k) -\beta H_B(k,j)$ in the three-site basis
$\{\psi_q\}$.  $-\beta H_A$ and $-\beta H_B$ have the form of
Eq.~\eqref{anis_eq:15}, with interaction constants
$\{t_A,$\,$J_A,$\,$V_A,$\,$\mu_A,$\,$\nu_A,$\,$G_A\}$ and
$\{t_B,$\,$J_B,$\,$V_B,$\,$\mu_B,$\,$\nu_B,$\,$G_B\}$ respectively.
The resulting matrix elements are listed in Table~\ref{anis_tab:3}.
We exponentiate the matrix blocks to find the elements $\langle
\psi_q|e^{-\beta H_A(i,k)-\beta H_B(k,j)}|\psi_{\bar{q}}\rangle$ which
enter on the right-hand side of Eq.~\eqref{anis_eq:13}.  In this way
the $\gamma_p$ are obtained as functions of the interaction constants
in the unrenormalized two-site Hamiltonians, $\gamma_p =
\gamma_p(\{t_A,J_A,\ldots\},\{t_B,J_B,\ldots\})$.

\begin{table}[t]
\centering {
\begin{tabular}{|c|c|c|c|c|}
 \hline
  $n$ & $p$ & $s$ & $m_s$ & Two-site basis states\\
  \hline
  $0$ & $+$ & $0$ & $0$ &$|\phi_{1}\rangle=|\circ\circ\rangle$ \\
  \hline
  $1$ & $+$ & $1/2$ & $1/2$ &$|\phi_{2}\rangle=\frac{1}{\sqrt{2}}\{|\uparrow
  \circ\rangle+|\circ\uparrow\rangle\}$\\ \hline
  $1$ & $-$ & $1/2$ & $1/2$ &$|\phi_{4}\rangle=\frac{1}{\sqrt{2}}\{|\uparrow
  \circ\rangle-|\circ\uparrow\rangle\}$\\ \hline
  $2$ & $-$ & $0$ & $0$ & $|\phi_{6}\rangle=\frac{1}{\sqrt{2}}\{|\uparrow\downarrow\rangle
  -|\downarrow\uparrow\rangle\}$\\ \hline
  $2$ & $+$ & $1$ & $1$ &
  $|\phi_{7}\rangle=|\uparrow\uparrow\rangle$\\ \hline
  $2$ & $+$ & $1$ & $0$ &
  $|\phi_{9}\rangle=\frac{1}{\sqrt{2}}\{|\uparrow\downarrow\rangle+|\downarrow
  \uparrow\rangle\}$\\ \hline
\end{tabular}}
\caption[Two-site basis states]{The two-site basis states, with the corresponding particle
number ($n$), parity ($p$), total spin ($s$), and total spin
$z$-component ($m_s$) quantum numbers.  The states
$|\phi_{3}\rangle$, $|\phi_{5}\rangle$, and $|\phi_{8}\rangle$ are
obtained by spin reversal from $|\phi_{2}\rangle$,
$|\phi_{4}\rangle$, and $|\phi_{7}\rangle$, respectively.}\label{anis_tab:1}
\end{table}

\begin{table}[t]
\centering {
\begin{tabular}{|c|c|c|c|c|}
 \hline
  $n$ & $p$ & $s$ & $m_s$ & Three-site basis states\\
  \hline
  $0$ & $+$ & $0$ & $0$ &$|\psi_{1}\rangle=|\circ\circ\,\circ\rangle$ \\
  \hline
  $1$ & $+$ & $1/2$ & $1/2$ &$|\psi_{2}\rangle=|\circ
  \uparrow
  \circ\rangle,\: |\psi_{3}\rangle=\frac{1}{\sqrt{2}}\{|\uparrow
  \circ\,\circ\rangle+|\circ\,\circ\uparrow\rangle\}$\\ \hline
  $1$ & $-$ & $1/2$ & $1/2$ &$|\psi_{6}\rangle=\frac{1}{\sqrt{2}}\{|\uparrow
  \circ\,\circ\rangle-|\circ\,\circ\uparrow\rangle\}$\\ \hline
$2$ & $+$ & $0$ & $0$ &
  $|\psi_{8}\rangle=\frac{1}{2}\{|\uparrow\downarrow\circ\rangle-
  |\downarrow\uparrow\circ\rangle-|\circ\uparrow\downarrow\rangle+
  |\circ\downarrow\uparrow\rangle\}$\\ \hline
   $2$ & $-$ & $0$ & $0$ &
  $|\psi_{9}\rangle=\frac{1}{2}\{|\uparrow\downarrow\circ\rangle-
  |\downarrow\uparrow\circ\rangle+|\circ\uparrow\downarrow\rangle-
  |\circ\downarrow\uparrow\rangle\},$\\
  &&&&$|\psi_{10}\rangle=\frac{1}{\sqrt{2}}\{|\uparrow\circ\downarrow\rangle-|\downarrow\circ\uparrow
  \rangle\}$\\\hline
  $2$ & $+$ & $1$ & $1$ &
  $|\psi_{11}\rangle=|\uparrow\circ\uparrow\rangle,\:
  |\psi_{12}\rangle=\frac{1}{\sqrt{2}}\{|\uparrow\uparrow\circ\rangle+|\circ\uparrow\uparrow
  \rangle\}$\\ \hline
  $2$ & $+$ & $1$ & $0$ &
  $|\psi_{13}\rangle=\frac{1}{2}\{|\uparrow\downarrow\circ\rangle+
  |\downarrow\uparrow\circ\rangle+|\circ\uparrow\downarrow\rangle+
  |\circ\downarrow\uparrow\rangle\},$\\
  &&&& $|\psi_{14}\rangle=\frac{1}{\sqrt{2}}
  \{|\uparrow\circ\downarrow\rangle+|\downarrow\circ\uparrow
  \rangle\}$\\ \hline
  $2$ & $-$ & $1$ & $1$ &
  $|\psi_{17}\rangle=\frac{1}{\sqrt{2}}\{|\uparrow\uparrow\circ\rangle-|\circ\uparrow\uparrow
  \rangle\}$\\ \hline
  $2$ & $-$ & $1$ & $0$ &
  $|\psi_{18}\rangle=\frac{1}{2}\{|\uparrow\downarrow\circ\rangle+
  |\downarrow\uparrow\circ\rangle-|\circ\uparrow\downarrow\rangle-
  |\circ\downarrow\uparrow\rangle\}$\\ \hline
  $3$ & $+$ & $1/2$ & $1/2$ &
  $|\psi_{20}\rangle=\frac{1}{\sqrt{6}}\{2|\uparrow\downarrow\uparrow\rangle-|\uparrow\uparrow
  \downarrow\rangle-|\downarrow\uparrow\uparrow\rangle\}$\\
  \hline
  $3$ & $-$ & $1/2$ & $1/2$ &
  $|\psi_{22}\rangle=\frac{1}{\sqrt{2}}\{|\uparrow\uparrow
  \downarrow\rangle-|\downarrow\uparrow\uparrow\rangle\}$\\
  \hline
  $3$ & $+$ & $3/2$ & $3/2$ &
  $|\psi_{24}\rangle=|\uparrow\uparrow\uparrow\rangle$ \\ \hline
  $3$ & $+$ & $3/2$ & $1/2$ &
  $|\psi_{25}\rangle=\frac{1}{\sqrt{3}}\{|\uparrow\downarrow\uparrow\rangle+|\uparrow\uparrow
  \downarrow\rangle+|\downarrow\uparrow\uparrow\rangle\}$ \\ \hline
\end{tabular}}
\caption[Three-site basis states]{The three-site basis states, with the corresponding
particle number ($n$), parity ($p$), total spin ($s$), and total
spin $z$-component ($m_s$) quantum numbers. The states
$|\psi_{4-5}\rangle$, $|\psi_{7}\rangle$, $|\psi_{15-16}\rangle$,
$|\psi_{19}\rangle$, $|\psi_{21}\rangle$, $|\psi_{23}\rangle$,
$|\psi_{26-27}\rangle$ are obtained by spin reversal from
$|\psi_{2-3}\rangle$, $|\psi_{6}\rangle$, $|\psi_{11-12}\rangle$,
$|\psi_{17}\rangle$, $|\psi_{20}\rangle$, $|\psi_{22}\rangle$,
$|\psi_{24-25}\rangle$, respectively.}\label{anis_tab:2}
\end{table}

\begingroup
\begin{table}
{\scriptsize
\begin{gather*}
\begin{array}{|c||c|}\hline
 & \psi_{1}\\
\hhline{|=#=|} \psi_{1} & 0\\ \hline
\end{array}\\
\begin{array}{|c||c|c|c|}\hline
 & \psi_{2} & \psi_{3} & \psi_{6}\\
\hhline{|=#=|=|=|} \psi_{2} & \parbox{0.5in}{$\mu_A + \mu_B - \nu_A + \nu_B$} & -\frac{1}{\sqrt{2}}(t_A+t_B) & \frac{1}{\sqrt{2}}(t_B-t_A)\\
\hline \psi_{3} & -\frac{1}{\sqrt{2}}(t_A+t_B) & \parbox{0.7in}{$\frac{1}{2}(\mu_A+\mu_B+\nu_A-\nu_B)$} & \parbox{0.7in}{$\frac{1}{2}(\mu_A - \mu_B + \nu_A + \nu_B)$}\\
\hline \psi_{6} & \frac{1}{\sqrt{2}}(t_B-t_A) & \parbox{0.7in}{$\frac{1}{2}(\mu_A - \mu_B + \nu_A + \nu_B)$} & \parbox{0.7in}{$\frac{1}{2}(\mu_A + \mu_B + \nu_A - \nu_B)$}\\
\hline
\end{array}\\
\begin{array}{|c||c|c|c|}\hline
 & \psi_{8} & \psi_{9} & \psi_{10}\\
\hhline{|=#=|=|=|} \psi_{8} &
\parbox{0.9in}{$\frac{1}{2}(\frac{3}{4}J_A + \frac{3}{4}J_B + V_A + V_B + 3\mu_A
+ 3\mu_B - \nu_A + \nu_B)$} &
\parbox{0.9in}{$\frac{1}{2}(\frac{3}{4}J_A - \frac{3}{4}J_B + V_A - V_B +
\mu_A - \mu_B + \nu_A + \nu_B)$} & \frac{1}{\sqrt{2}}(t_A-t_B)\\
\hline \psi_{9} & \parbox{0.9in}{$\frac{1}{2}(\frac{3}{4}J_A -
\frac{3}{4}J_B + V_A - V_B + \mu_A - \mu_B + \nu_A + \nu_B)$} &
\parbox{0.9in}{$\frac{1}{2}(\frac{3}{4}J_A + \frac{3}{4}J_B + V_A + V_B +
3\mu_A
+ 3\mu_B - \nu_A + \nu_B)$}  & -\frac{1}{\sqrt{2}}(t_A+t_B)\\
\hline  \psi_{10} & \frac{1}{\sqrt{2}}(t_A-t_B) &
   -\frac{1}{\sqrt{2}} (t_A + t_B)  & \parbox{0.5in}{$\mu_A+\mu_B+\nu_A-\nu_B$}\\
\hline
\end{array}\\
\begin{array}{|c||c|c|c|}\hline
 & \psi_{11} & \psi_{12} & \psi_{17}\\
\hhline{|=#=|=|=|} \psi_{11} & \parbox{0.5in}{$\mu_A + \mu_B +
\nu_A - \nu_B$} & -\frac{1}{\sqrt{2}}(t_A+t_B) & \frac{1}{\sqrt{2}}(t_A-t_B) \\
\hline \psi_{12} & -\frac{1}{\sqrt{2}}(t_A+t_B) & \parbox{0.9in}{$\frac{1}{2}(-\frac{1}{4}J_A - \frac{1}{4}J_B + V_A + V_B + 3\mu_A + 3\mu_B - \nu_A + \nu_B)$} & \parbox{0.9in}{$\frac{1}{2}(-\frac{1}{4}J_A + \frac{1}{4}J_B + V_A - V_B + \mu_A - \mu_B + \nu_A + \nu_B)$}\\
\hline  \psi_{17} & \frac{1}{\sqrt{2}}(t_A-t_B) &
   \parbox{0.9in}{$\frac{1}{2}(-\frac{1}{4}J_A + \frac{1}{4}J_B + V_A - V_B + \mu_A - \mu_B + \nu_A + \nu_B)$}  & \parbox{0.9in}{$\frac{1}{2}(-\frac{1}{4}J_A - \frac{1}{4}J_B + V_A + V_B + 3\mu_A + 3\mu_B - \nu_A + \nu_B)$}  \\
\hline
\end{array}\\
\begin{array}{|c||c|c|c|}\hline
 & \psi_{13} & \psi_{14} & \psi_{18}\\
\hhline{|=#=|=|=|} \psi_{13} &
\parbox{0.9in}{$\frac{1}{2}(-\frac{1}{4}J_A - \frac{1}{4}J_B + V_A + V_B + 3\mu_A
+ 3\mu_B - \nu_A + \nu_B)$} &
-\frac{1}{\sqrt{2}}(t_A+t_B) &  \parbox{0.9in}{$\frac{1}{2}(-\frac{1}{4}J_A + \frac{1}{4}J_B + V_A - V_B + \mu_A - \mu_B + \nu_A + \nu_B)$}\\
\hline \psi_{14} & -\frac{1}{\sqrt{2}}(t_A+t_B) &
\parbox{0.5in}{$\mu_A + \mu_B +
\nu_A - \nu_B$} & \frac{1}{\sqrt{2}}(t_A-t_B)\\
\hline  \psi_{18} & \parbox{0.9in}{$\frac{1}{2}(-\frac{1}{4}J_A +
\frac{1}{4}J_B + V_A - V_B + \mu_A - \mu_B + \nu_A + \nu_B)$} &
   \frac{1}{\sqrt{2}}(t_A-t_B) & \parbox{0.9in}{$\frac{1}{2}(-\frac{1}{4}J_A - \frac{1}{4}J_B + V_A + V_B + 3\mu_A
+ 3\mu_B - \nu_A + \nu_B)$}  \\
\hline
\end{array}\\
\begin{array}{|c||c|c|}\hline
 & \psi_{20} & \psi_{22}\\
\hhline{|=#=|=|} \psi_{20} & \parbox{1in}{$\frac{1}{2}J_A+\frac{1}{2}J_B+V_A+V_B+2\mu_A+2\mu_B$} & \frac{\sqrt{3}}{4}(J_B-J_A)\\
\hline \psi_{22} & \frac{\sqrt{3}}{4}(J_B-J_A) &
\parbox{0.9in}{$V_A+V_B+2\mu_A+2\mu_B$}\\
\hline
\end{array}\\
\begin{array}{|c||c|}\hline
 & \psi_{24}\\
\hhline{|=#=|} \psi_{24} & \parbox{1in}{$-\frac{1}{4}J_A-\frac{1}{4}J_B+V_A+V_B+2\mu_A+2\mu_B$} \\
\hline
\end{array}\quad
\begin{array}{|c||c|}\hline
 & \psi_{25}\\
\hhline{|=#=|} \psi_{25}
&\parbox{1in}{$-\frac{1}{4}J_A-\frac{1}{4}J_B+V_A+V_B+2\mu_A+2\mu_B$}
\\ \hline
\end{array}
\end{gather*}}
\caption[Diagonal matrix blocks of the unrenormalized three-site
Hamiltonian $-\beta H_A(i,k)-\beta H_B(k,j)$]{Diagonal matrix blocks
of the unrenormalized three-site Hamiltonian $-\beta H_A(i,k)-\beta
H_B(k,j)$.  The Hamiltonian being invariant under spin-reversal, the
spin-flipped matrix elements are not shown.  The additive constant
contribution $G_A+G_B$, occurring at the diagonal terms, is also not
shown.}\label{anis_tab:3}
\end{table}
\endgroup

\begingroup
\begin{table}
{\scriptsize
\begin{gather*}
\begin{array}{|c||c|c|c|c|c|c|} \hline & \parbox{0.08in}{$\phi_{1}$} &\phi_{2} &\phi_{4} & \phi_{6} &
\phi_{7} &\phi_{9}\\
\hhline{|=#=|=|=|=|=|=|} \parbox{0.1in}{$\phi_{1}$} & \parbox{0.08in}{$G^\prime$} & \multicolumn{5}{c|}{}\\
\cline{1-4} \parbox{0.1in}{$\phi_{2}$} && \parbox{0.4in}{\centering $-t^\prime+\mu^\prime+G^\prime$} & \nu^\prime & \multicolumn{3}{c|}{0}\\
\cline{1-1}\cline{3-4} \parbox{0.1in}{$\phi_{4}$} &
\multicolumn{1}{c|}{} & \nu^\prime
&\parbox{0.4in}{\centering $t^\prime + \mu^\prime+G^\prime$} &\multicolumn{3}{c|}{}\\
\cline{1-1}\cline{3-5} \parbox{0.1in}{$\phi_{6}$} &
\multicolumn{3}{c|}{} &
\multicolumn{1}{c|}{\parbox{0.575in}{\centering $\frac{3}{4}J^\prime
+V^\prime+2\mu^\prime+G^\prime$}} &
\multicolumn{2}{c|}{}\\
\cline{1-1}\cline{5-6} \parbox{0.1in}{$\phi_{7}$} &
\multicolumn{4}{c|}{0}
&\multicolumn{1}{c|}{\parbox{0.485in}{\centering $-\frac{1}{4}J^\prime+V^\prime+2\mu^\prime+G^\prime$}}&\\
\cline{1-1}\cline{6-7} \parbox{0.1in}{$\phi_{9}$} &
\multicolumn{5}{c|}{}
& \parbox{0.485in}{\centering $-\frac{1}{4}J^\prime+V^\prime+2\mu^\prime+G^\prime$}\\
\hline
\end{array}
\end{gather*}}
\caption[Block-diagonal matrix of the renormalized two-site
Hamiltonian\newline $-\beta^\prime H^\prime(i,j)$]{Block-diagonal matrix of
the renormalized two-site Hamiltonian $-\beta^\prime H^\prime(i,j)$.
The Hamiltonian being invariant under spin-reversal, the spin-flipped
matrix elements are not shown.}\label{anis_tab:4}
\end{table}
\endgroup

The matrix elements of $-\beta^{\prime }H_{A,B}^{\prime }(i,j)$ in
the $\{\phi_p\}$ basis are shown in Table~\ref{anis_tab:4}. Exponentiating this
matrix, we solve for the renormalized interaction constants
$(t^\prime$, $J^\prime$, $V^\prime$, $\mu^\prime$, $\nu^\prime$, $G^\prime)$ in
terms of the $\gamma_p$:
\begin{gather}
t^\prime = u,\qquad  J^\prime =
\ln\frac{\gamma_6}{\gamma_7},\nonumber\\
V^\prime = \frac{1}{4}\left\{\ln(\gamma_1^4 \gamma_6
\gamma_7^3)-8v\right\},\qquad  \mu^\prime = v - \ln
\gamma_1,\nonumber\\
\nu^\prime = \frac{2u\gamma_0}{\gamma_4-\gamma_2},\qquad
G^\prime=\ln\gamma_1,\label{anis_eq:16}
\end{gather}
where
\begin{gather*}
v = \frac{1}{2}\ln\left(\gamma_2\gamma_{4}-\gamma_0^2\right)\,,\\
u = \frac{\gamma_{4}-\gamma_2}
{\sqrt{\left(\gamma_{4}-\gamma_2\right)^2+4\gamma_0^2}} \cosh^{-1}
\left(\frac{\gamma_4+\gamma_2}{2e^v}\right).
\end{gather*}

The renormalization-group transformation described by
Eqs.~\eqref{anis_eq:13}-\eqref{anis_eq:16} can be expressed as a mapping of a
three-site Hamiltonian with bonds having interaction constants
$\mathbf{K}_A = \{t_A,J_A,V_A,\mu_A,\nu_A,G_A\}$ and $\mathbf{K}_B =
\{t_B,J_B,V_B,\mu_B,\nu_B,G_B\}$ onto a two-site Hamiltonian with
interaction constants
\begin{equation}\label{anis_eq:17}
\mathbf{K}^\prime = \mathbf{R}(\mathbf{K}_A,\mathbf{K}_B)\,.
\end{equation}
When $\nu_A = \nu_B = 0$, this mapping has the property that if
$\mathbf{R}(\mathbf{K}_A,\mathbf{K}_B) =
\{t^\prime,J^\prime,V^\prime,\mu^\prime,\nu^\prime,G^\prime\}$, then
$\mathbf{R}(\mathbf{K}_B,\mathbf{K}_A)$ gives the same result,
except that the sign of $\nu^\prime$ is switched.  So
$\mathbf{R}(\mathbf{K}_A,\mathbf{K}_A)$ has a zero $\nu^\prime$
component when $\nu_A = 0$.

>From Eq.~\eqref{anis_eq:9}, the renormalized $xy$- and $z$-bond
interaction constants are
\begin{equation}\label{anis_eq:18}\begin{split}
\mathbf{K}^\prime_{xy} &= 2
\mathbf{R}(\mathbf{K}_{xy},\mathbf{K}_{xy}) +
\mathbf{R}(\mathbf{K}_{xy},\mathbf{K}_{z}) +
\mathbf{R}(\mathbf{K}_{z},\mathbf{K}_{xy})\,,\\
\mathbf{K}^\prime_{z} &= \mathbf{R}(\mathbf{K}_{z},\mathbf{K}_{z}) +
\mathbf{R}(\mathbf{K}_{xy},\mathbf{K}_{z}) +
2\mathbf{R}(\mathbf{K}_{z},\mathbf{K}_{xy})\,.
\end{split}
\end{equation}
The staggered $\nu^\prime$ term cancels out in
$\mathbf{K}^\prime_{xy}$.  In constructing the anisotropic
hierarchical lattice, we could have used a graph in which the lowest
two bonds in Fig.~\ref{anis_fig:1}(b) are interchanged.  Averaging over these two
realizations,
\begin{equation}\label{anis_eq:19}\begin{split}
\mathbf{K}^\prime_{z} &= \mathbf{R}(\mathbf{K}_{z},\mathbf{K}_{z}) +
\frac{3}{2}\mathbf{R}(\mathbf{K}_{xy},\mathbf{K}_{z}) +
\frac{3}{2}\mathbf{R}(\mathbf{K}_{z},\mathbf{K}_{xy})\,,
\end{split}
\end{equation}
the $\nu^\prime$ term cancels out in $\mathbf{K}^\prime_z$ as well.

\section[Phase Diagrams and Expectation Values as a Function of\\ Anisotropy]{Phase Diagrams and Expectation Values as a Function of Anisotropy}

Thermodynamic properties of the system, including the global phase
diagram and expectation values of operators occurring in the
Hamiltonian, are obtained from the analysis of the
renormalization-group flows~\cite{ABOP}. The initial conditions for
the flows are the interaction constants in the original anisotropic
$tJ$ Hamiltonian.  For the numerical results presented below, we use
the following initial form: $t_{xy} = t$, $t_z = \alpha_t t$,
$J_{xy} = J$, $J_z = \alpha_J J$, $V_{xy} = J_{xy}/4$, $V_{z} =
J_z/4$, where $0 \le \alpha_t, \alpha_J \le 1$. For the anisotropy
parameters $\alpha_t$ and $\alpha_J$, we use $\alpha_J =
\alpha_t^2$, as dictated from the derivation of the $tJ$ Hamiltonian
from the large-$U$ limit of the Hubbard model~\cite{AShankarSingh}.

\begin{figure*}
\centering
\includegraphics*[scale=0.95]{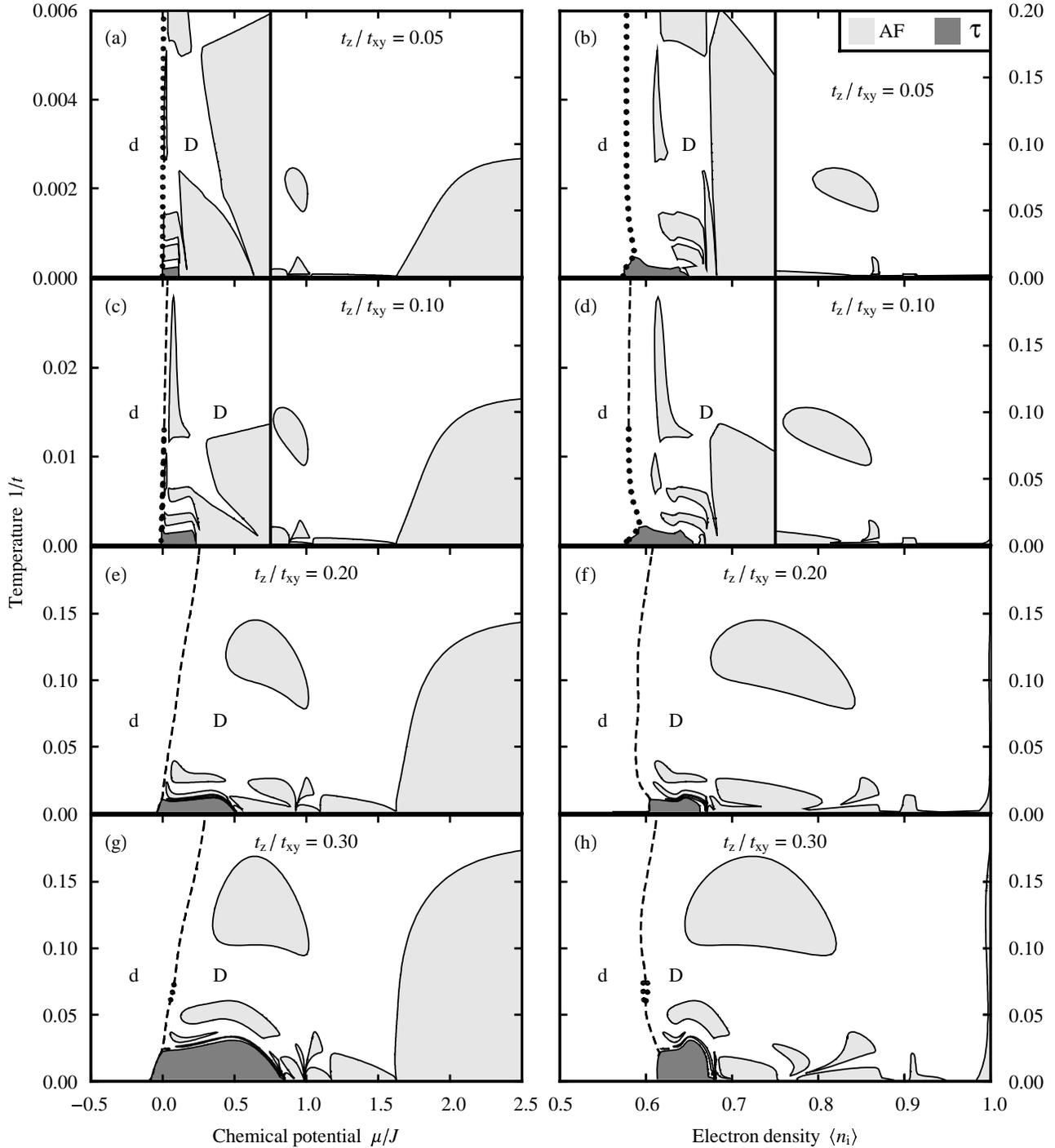}
\caption[Phase diagrams of the anisotropic $tJ$ model with $J/t =
0.444$ in temperature vs. chemical potential (first column) and
temperature vs. electron density (second column)]{Phase diagrams of
the anisotropic $tJ$ model with $J/t = 0.444$ in temperature
vs. chemical potential (first column) and temperature vs. electron
density (second column).  The degree of anisotropy varies from
$t_z/t_{xy} = 0.05$ in Fig.~\ref{anis_fig:2}(a)-(b) to $t_z/t_{xy} =
0.30$ in Fig.~\ref{anis_fig:2}(g)-(h).  Note the expanded temperature
scales on the left panels of Fig.~\ref{anis_fig:2}(a)-(d).  The dense
disordered (D), dilute disordered (d), antiferromagnetic (A), and
$\tau$ phases are shown. The A and $\tau$ regions are colored light
and dark gray respectively. Second-order phase transitions are drawn
with full curves, first-order transitions with dotted curves. The
unmarked areas within the dotted curves in the temperature
vs. electron density figures are narrow coexistence regions between
the two phases at either side.  Dashed curves are not phase
transitions, but disorder lines between the dense disordered and
dilute disordered phases.}\label{anis_fig:2}
\end{figure*}

\begin{figure*}
\centering
\includegraphics*[scale=0.95]{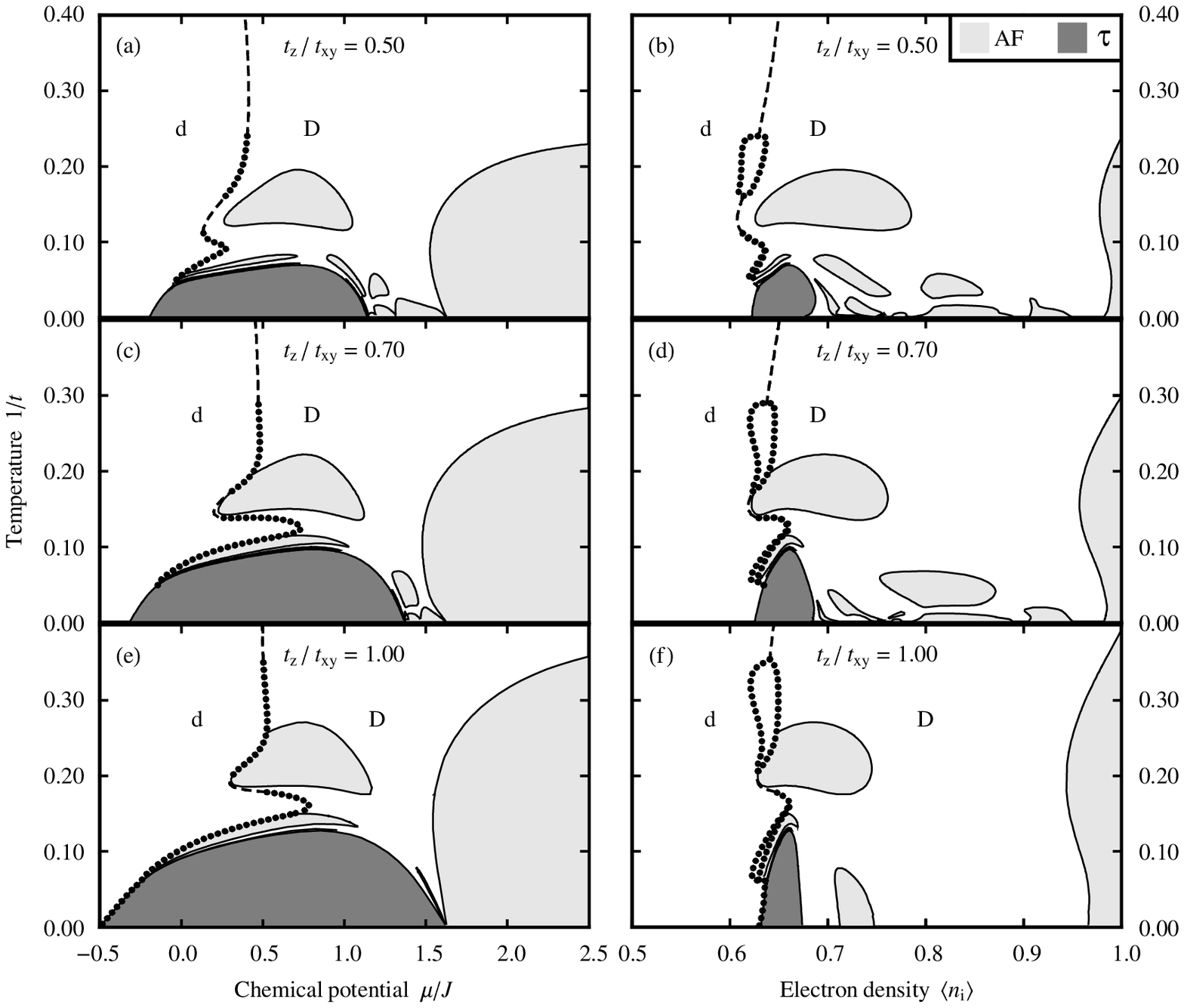}
\caption{The continuation of the phase diagrams in Fig.~\ref{anis_fig:2} for
$t_{z}/t_{xy}$ between 0.5 and 1.}\label{anis_fig:3}
\end{figure*}

Phase diagrams for the coupling $J/t = 0.444$ and various values of
$\alpha_t = t_z/t_{xy}$ are shown in Figs.~\ref{anis_fig:2} and
\ref{anis_fig:3}.  The temperature variable is $1/t$, and the diagrams
are plotted both in terms of chemical potential $\mu/J$ and electron
density $\langle n_i
\rangle$.  The phases in the diagrams are those found in earlier
studies of the isotropic $d=3$ $tJ$
model~\cite{AFalicovBerkerT,AFalicovBerker}, which can be consulted
for a more detailed description.  Here we summarize the salient
features of the phases.

\begin{figure*}
\centering
\includegraphics*[scale=0.97]{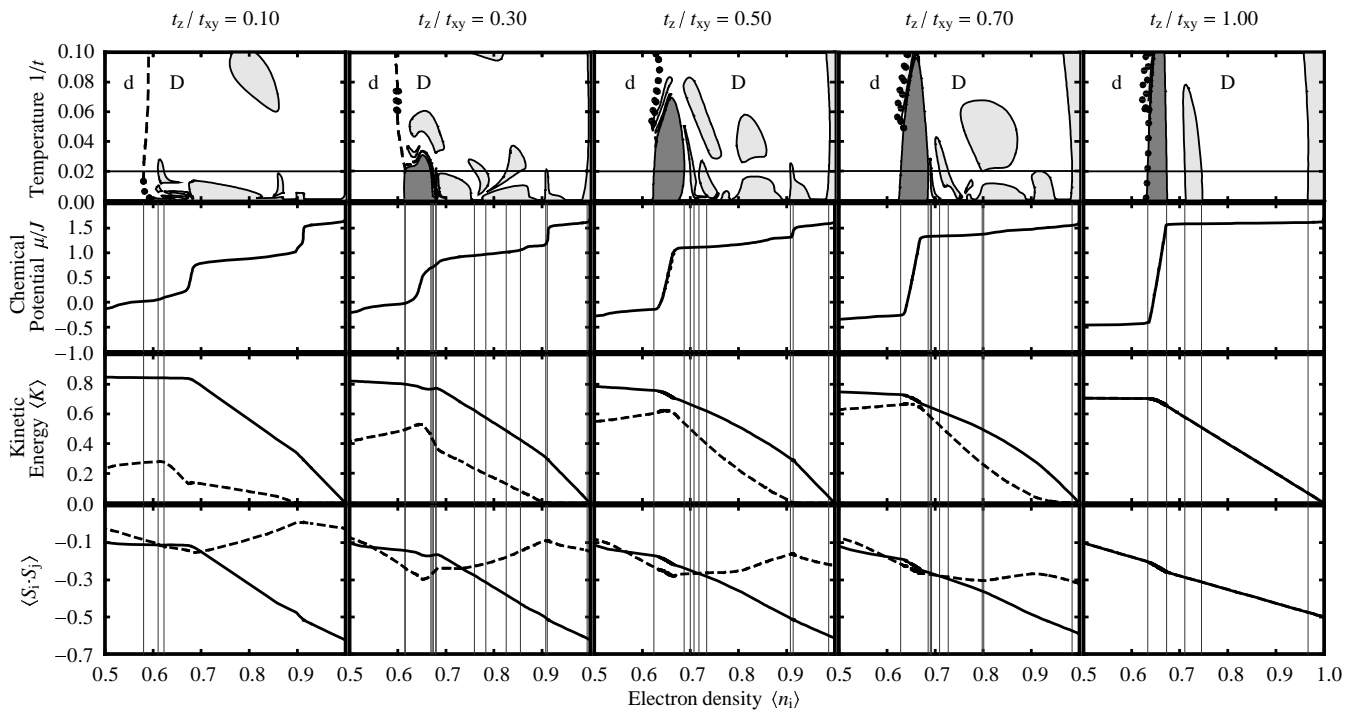}
\caption[Thermodynamic properties along slices of the phase diagrams
at the constant temperature $1/t = 0.02$]{Thermodynamic properties along slices of the phase diagrams
at the constant temperature $1/t = 0.02$. The degree of anisotropy
varies from $t_z/t_{xy} = 0.10$ in the first column to $t_z/t_{xy} =
1.00$ in the last column. The top row contains the temperature vs.
electron density phase diagrams and a thin horizontal line marking
the slice. The antiferromagnetic and $\tau$ phases are colored light
and dark gray respectively.  The rows below this show the chemical
potential $\mu/J$, kinetic energy $\langle K \rangle = -\sum_\sigma
\langle c^\dagger_{i\sigma}c_{j\sigma} +
c^\dagger_{j\sigma}c_{i\sigma}\rangle$, and nearest-neighbor
spin-spin correlation $\langle \mathbf{S}_i \cdot
\mathbf{S}_j\rangle$.  For the $\langle K \rangle$ and $\langle
\mathbf{S}_i \cdot \mathbf{S}_j\rangle$ graphs, full curves denote
results for nearest neighbors along the $xy$ plane, while dashed
curves denote those for nearest neighbors along the $z$ direction.
(In the $t_z/t_{xy} = 1$ column, these two curves overlap.)  Thin
vertical lines mark the location of phase transitions.}\label{anis_fig:4}
\end{figure*}

Each phase is associated with a completely stable fixed point (sink)
of the renormalization-group flows, and thermodynamic densities
calculated at the fixed point epitomize (and
determine~\cite{AHinczBerker2}, e.g., as seen in the results
displayed in Fig.~\ref{anis_fig:4}) characteristics of the entire
phase. The results are shown in Table~\ref{anis_tab:5}.  The {\bf
dilute disordered (d)} and {\bf dense disordered (D)} phases have
$\langle n_i \rangle = 0$ and 1 at their respective phase sinks, so
the electron densities in these phases are accordingly small in the
one case and close to 1 in the other.  Both phases lack long-range
spin order, since $\langle \mathbf{S}_i \cdot \mathbf{S}_j \rangle =
0$ at the sinks.  On the other hand, the {\bf antiferromagnetic (A)}
phase has $\langle n_i \rangle = 1$ and a nonzero nearest-neighbor
spin-spin correlation $\langle \mathbf{S}_i \cdot \mathbf{S}_j
\rangle = 1/4$ at the phase sink. Since nearest-neighbor spins at
the sink are distant members of the same sublattice in the
unrenormalized system, this positive value for $\langle \mathbf{S}_i
\cdot \mathbf{S}_j \rangle$ is expected, and leads to $\langle
\mathbf{S}_i \cdot \mathbf{S}_j \rangle < 0$ for nearest neighbors
of the original system, as seen in the last row of
Fig.~\ref{anis_fig:4}.

\begin{table}[t]
\centering
\begin{tabular}{|c|c|c|c|c|}
 \hline
  Phase sink & \multicolumn{4}{c|}{Expectation values}\\
  \cline{2-5}
  & $-\sum_\sigma \langle c^\dagger_{i\sigma}c_{j\sigma} +
c^\dagger_{j\sigma}c_{i\sigma}\rangle$  & $\langle n_i \rangle$ &
$\langle \mathbf{S}_i \cdot
\mathbf{S}_j \rangle$ & $\langle n_i n_j \rangle$\\
  \hline
  d & 0 & 0 & 0 & 0\\ \hline
  D & 0 & 1 & 0 & 1\\ \hline
  A & 0 & 1 & $\frac{1}{4}$ & 1 \\\hline
  $\tau$ & $\frac{2}{3}$ & $\frac{2}{3}$ & $-\frac{1}{4}$ & $\frac{1}{3}$ \\\hline
\end{tabular}
\caption{Expectation values at the phase-sink fixed points.}\label{anis_tab:5}
\end{table}

In the antiferromagnetic and the two disordered phases, the electron
hopping strengths $t_{xy}$ and $t_{z}$ tend to zero after repeated
rescalings.  The system is either completely empty or filled in this
limit, and the expectation value of the kinetic energy operator
$\langle K \rangle \equiv - \sum_\sigma \langle
c^\dagger_{i\sigma}c_{j\sigma} +
c^\dagger_{j\sigma}c_{i\sigma}\rangle$ is zero at the sink. The {\bf
$\pmb\tau$ phase} is interesting in contrast because the magnitudes
of $t_{xy}$ and $t_z$ both tend to $\infty$, and we find partial
filling, $\langle n_i \rangle = 2/3$, and a nonzero kinetic energy
$\langle K \rangle = 2/3$ at the phase sink. It should be
recalled that we have shown in a previous work~\cite{AHinczBerker2}
that the superfluid weight has a pronounced peak in the $\tau$
phase, there is evidence of a gap in the quasiparticle spectrum, and
the free carrier density in the vicinity of the $\tau$ phase has
properties seen experimentally in cuprates~\cite{ABernhard2,APuchkov}.

Figs.~\ref{anis_fig:2} and \ref{anis_fig:3} clearly demonstrate that
the $\tau$ phase is not unique to the isotropic $d=3$ case, but exists
at all values of $t_z/t_{xy}$, even persisting in the weak interplane
coupling limit.  Fig.~\ref{anis_fig:2} shows the evolution of the
phase diagram in the strongly anisotropic regime, for $t_z/t_{xy}$
between 0.05 and 0.30, while Fig.~\ref{anis_fig:3} completes the
evolution from $t_z/t_{xy} = 0.5$ to the fully isotropic case where
$t_z/t_{xy} = 1$.  The $\tau$ phase is present even for $t_z/t_{xy} =
0.05$ and 0.10, but only at very low temperatures close to the d/D
first-order phase transition that itself is distinct by its very
narrow coexistence region.  As the interplane coupling is increased,
the $\tau$ phase transition temperatures also get larger, but the
density range in which the phase occurs, namely $\langle n_i \rangle$
around 0.65, remains unchanged.

As expected, the antiferromagnetic transition temperatures also
increase with the interplane coupling. The phase diagrams all share
an antiferromagnetic region near $\langle n_i \rangle=1$, which is
confined to $\langle n_i \rangle$ very close to 1 in the strongly
anisotropic limit, but becomes more stable to hole doping as
$t_z/t_{xy}$ gets larger.  Away from $\langle n_i\rangle=1$, in the
range of 5-35\% hole doping, there are thin slivers and islands of
antiferromagnetism separated by regions of the dense disordered
phase.  For $t_z/t_{xy} = 1$, we see these mostly around the $\tau$
phase, but as anisotropy is introduced into the system, the
structure of the antiferromagnetic regions becomes more complex, and
spread out over a wider range of densities.  The lamellar structure
of A and D phases here potentially indicates an underlying
incommensurate order~\cite{AFalicovBerkerT}. The physical
significance of this possibility will be discussed below.

Further insight into the nature of the $\tau$ phase can be gained by
looking at thermodynamic densities on a constant-temperature slice of
the phase diagram.  Fig.~\ref{anis_fig:4} plots the chemical potential
$\mu/J$, kinetic energy $\langle K \rangle$, and nearest-neighbor
spin-spin correlation $\langle \mathbf{S}_i\cdot \mathbf{S}_j\rangle$
at the temperature $1/t = 0.02$ for several values of $t_z/t_{xy}$.
Averages over the $xy$ bonds, $\langle \: \rangle_{xy}$ are drawn with
full curves in the figure, and averages taken over the $z$ bonds,
$\langle \: \rangle_z$ are drawn with dashed curves.

Consider first the kinetic energy expectation value $\langle K
\rangle = -\sum_\sigma \langle c^\dagger_{i\sigma}c_{j\sigma} +
c^\dagger_{j\sigma}c_{i\sigma}\rangle$.  The $xy$ bond kinetic
energy $\langle K \rangle_{xy}$ grows with hole doping until the
density range where the $\tau$ phase occurs, and then levels off.
This behavior is seen for the whole range of $t_z/t_{xy}$.  We
can compare our calculational result here with experimental results
in cuprates, by relating the kinetic energy expectation value in the
$tJ$ model to the density of free carriers as
follows~\cite{AHinczBerker2}. $\langle K \rangle$ and the total
weight of $\sigma_1(\omega,T)$, the real part of the optical
conductivity, satisfy the sum rule~\cite{ATan}
\begin{equation}\label{tc_eq:8}
\int_0^\infty d\omega\, \sigma_1(\omega,T) = \frac{\pi e^2}{2}
\langle K \rangle\,.
\end{equation}
To understand this sum rule, we keep in mind that the $tJ$
Hamiltonian describes a one-band system, so cannot account for
interband transitions. For real materials, the full conductivity sum
rule has the form
\begin{equation}\label{tc_eq:10}
\int_0^\infty d\omega\, \sigma_1(\omega,T) = \frac{\pi e^2
n}{2m}\,,
\end{equation}
where $n$ is the total density of electrons and $m$ is the free
electron mass.  The right-hand side of Eq.~\eqref{tc_eq:10} is
independent of electron-electron interactions, in contrast to the
right-hand side of Eq.~\eqref{tc_eq:8}, where $\langle K\rangle$
varies with the interaction strengths in the Hamiltonian.  The
optical conductivity of actual materials incorporates both
transitions within the conduction band and those to higher bands,
while the $tJ$ model contains only the conduction band.  We can look
at Eq.~\eqref{tc_eq:8} as a partial sum rule~\cite{ABaeriswyl,ATan},
which reflects the spectral weight of the free carriers in the
conduction band.

The experimental quantity we are interested in is the
density of free carriers, which in actual materials is
calculated from the low-frequency spectral
weight~\cite{AOrenstein},
\begin{equation}\label{tc_eq:10bb}
n_{\text{free}}(T) = \frac{2m_b}{\pi e^2} \int_0^{\omega_0}
d\omega\,\sigma_1(\omega,T)\,,
\end{equation}
where $m_b$ is the effective band mass of the electrons.  For
cuprates, the cut-off frequency is typically chosen around $\hbar
\omega_0 \approx 1$ eV so as to include only intraband transitions.
In comparison with the $tJ$ model, we identify the right-hand side
of Eq.~\eqref{tc_eq:8} with
\begin{equation}\label{tc_eq:10b}
\frac{\pi e^2}{2} \langle K \rangle = \frac{\pi e^2
n_{\text{free}}(T)}{2m_b}\,.
\end{equation}

Puchkov {\it et al.} ~\cite{APuchkov} have studied the in-plane
optical conductivity of a variety of cuprates, and found that the
low-frequency spectral weight increases with doping until the doping
level optimal for superconductivity is reached, and then remains
approximately constant in the overdoped regime.  This behavior of
$n_\text{free}/m_b$ is qualitatively reproduced in our results for
$\langle K \rangle_{xy}$.

As for $\langle K \rangle_z$, it is significantly reduced with
increasing anisotropy, since interplane hopping is suppressed.
$\langle K \rangle_z$ peaks in the $\tau$ phase, and decreases for
larger dopings.  This small peak in $\langle K \rangle_z$, which is
most pronounced in the strongly anisotropic regime, is accompanied by
an enhancement in the $\tau$ phase of the $z$-bond antiferromagnetic
nearest-neighbor spin-spin correlation, $\langle \mathbf{S}_{i} \cdot
\mathbf{S}_{j} \rangle_z$.  For the $xy$ planes, $\langle
\mathbf{S}_{i} \cdot \mathbf{S}_{j} \rangle_{xy}$ generally increases
(i.e., becomes less negative) with hole doping from a large negative
value near $\langle n_i \rangle=1$, as additional holes weaken the
antiferromagnetic order.  This increase becomes much less pronounced
when the $\tau$ phase is reached, and $\langle \mathbf{S}_{i} \cdot
\mathbf{S}_{j} \rangle_{xy}$ becomes nearly constant for large hole
dopings in the strongly anisotropic limit.  Rather than increasing
with hole doping, $\langle \mathbf{S}_{i} \cdot \mathbf{S}_{j}
\rangle_{z}$ shows the opposite behavior in the 10-35\% doping range,
decreasing and reaching a minimum within the $\tau$ phase.

The final aspect of the $\tau$ phase worth noting is the large
change in chemical potential $\mu/J$ over the narrow density range
where this phase occurs.  This is in contrast to broad regions at
smaller hole dopings where the chemical potential change is much
shallower, and which correspond to those parts of the phase diagram
where A and D alternate.  We can see this directly in the phase
diagram topology in Figs.~\ref{anis_fig:2} and \ref{anis_fig:3}, particularly for larger
$t_z/t_{xy}$.  The $\tau$ phase has a very wide extent in terms of
chemical potential, but becomes very narrow in the corresponding
electron density diagram.  The converse is true for the complex
lamellar structure of A and D phases sandwiched between the $\tau$
phase and the main antiferromagnetic region near $\langle n_i
\rangle=1$.  We shall return to this point in our discussion of the
purely two-dimensional results.

One can compare our phase diagram results for the $tJ$ model in
the strongly anisotropic limit to the large body of work done on the
square-lattice $tJ$ model.  Here a primary focus has been on the
possibility of a superconducting ground-state (or other types of
order) away from half-filling, with the presumption that a
zero-temperature long-range ordered state in the two-dimensional
system would develop a finite transition temperature with the
addition of interplanar coupling.  Numerical studies using exact
diagonalization of finite clusters and variational calculations with
trial ground-state wavefunctions have shown enhanced $d_{x^2-y^2}$
pair-pair correlation for $J/t \sim 3$ near $\langle n \rangle =
1/2$~\cite{DagottoRiera,DagottoRiera2}, and variational approaches
have yielded indications of $d$-wave superconductivity for more
realistic parameters like $J/t = 0.4-0.5$ over a range of densities
$0.6 < \langle n_i \rangle < 1$~\cite{Kohno,YokoyamaOgata,Sorella}.
Slave-boson mean-field theory of the $tJ$ model has also predicted a
phase diagram with a $d$-wave superconducting phase within this same
doping range away from half-filling~\cite{LeeReview}.  The least
biased approach, through high-temperature series expansions, has
given mixed signals on this issue.  Pryadko {\it et al.}
~\cite{PryadkoKivelsonZachar}, using a series through ninth order in
inverse temperature, did not observe an increase in the $d$-wave
superconducting susceptibility for the doped system at low
temperatures for $J/t < 1$.  On the other hand, Koretsune and
Ogata~\cite{KoretsuneOgata}, using a series up to twelfth order, did
see a rapid rise in the correlation length for $d$-wave pairing with
decreasing temperature for densities $0.5 < \langle n_i \rangle <
0.9$, with the largest correlations around $\langle n_i \rangle \sim
0.6$.  A similar calculation by Puttika and
Luchini~\cite{PutikkaLuchini} also gave a broad, growing peak in the
low-temperature $d$-wave correlation length, but with the maximum
shifted to smaller dopings around $\langle n_i \rangle \sim 0.75$.
Thus the fact that we see the $\tau$ phase emerge near these
densities for any non-zero interplanar coupling in the anisotropic
$tJ$ model, fits with prevailing evidence for an instability toward
$d$-wave superconductivity away from half-filling in the
two-dimensional system.

\section{The Two-Dimensional Isotropic $tJ$ Model and Chemical Potential Shift}

The above analysis leads to a basic question: how do results for a
strongly anisotropic $d=3$ $tJ$ model compare to results obtained
directly through a renormalization-group approach for the isotropic
$d=2$ system?  The latter was studied in
Refs.~\cite{AFalicovBerker,AFalicovBerkerT}, which yielded a phase
diagram with only dense and dilute disordered phases, separated by a
first-order transition at low temperatures, ending in a critical
point, but only for low values of $t/J$. The absence of any
antiferromagnetic order is consistent with the Mermin-Wagner
theorem~\cite{AMerminWagner}. As seen above, at least a weak coupling
in the $z$ direction is required for a finite N{\'e}el temperature.
What about the absence in $d=2$ of the $\tau$ phase? It turns out that
there is a pre-signature of the $\tau$ phase in $d=2$, and it appears
exactly where we find the actual phase upon adding the slightest
interplane coupling.

\begin{figure}
\centering \hspace{2.5pt}\includegraphics*[scale=1]{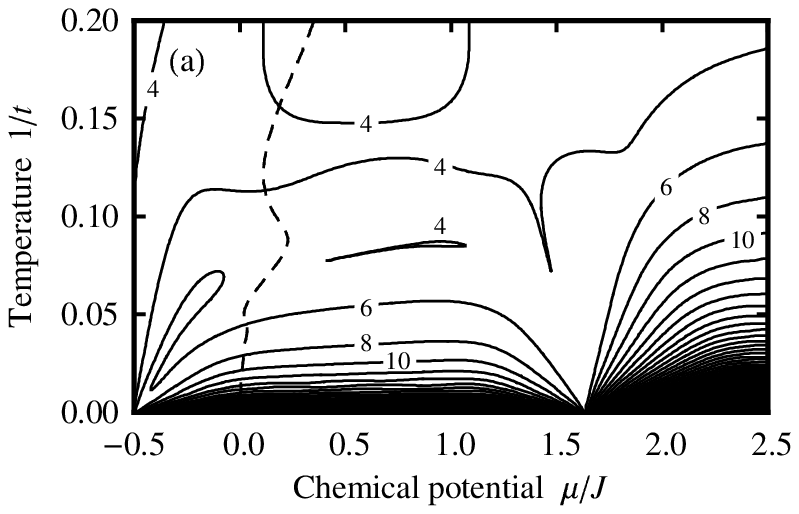}
\vspace{1em}

\includegraphics*[scale=1]{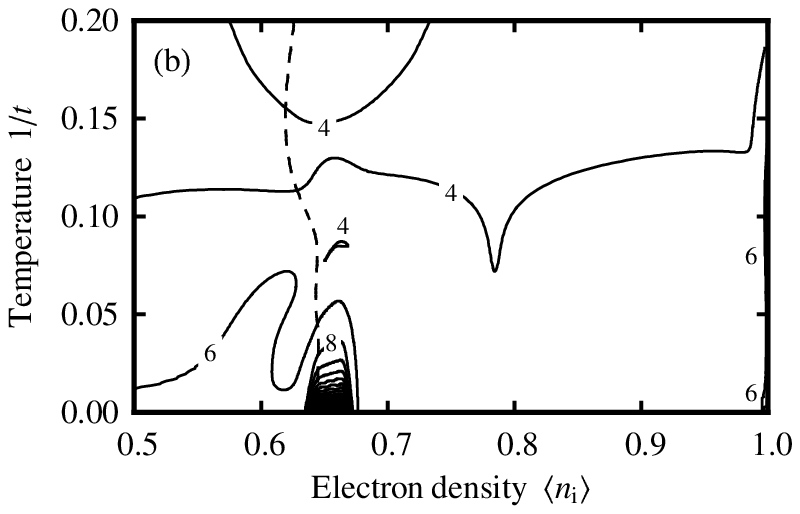}
\caption[Contour diagrams showing the number of iterations required
to reach a disordered phase sink in the $d=2$ \: $tJ$ model with
$J/t = 0.444$]{Contour diagrams showing the number of iterations
required to reach a disordered phase sink in the $d=2$ isotropic
$tJ$ model with $J/t = 0.444$. Fig.~\ref{anis_fig:5}(a) is plotted
in terms of temperature vs. chemical potential, while
Fig.~\ref{anis_fig:5}(b) is in terms of temperature vs. electron
density.  Note the accumulation of contours towards the $\tau$
ranges of the chemical potential and density.  The disorder line,
along which the trajectories eventually cross over from the $\tau$
region to disorder, is shown as dashed.}\label{anis_fig:5}
\end{figure}

\begin{figure}
\centering
\includegraphics*[scale=1]{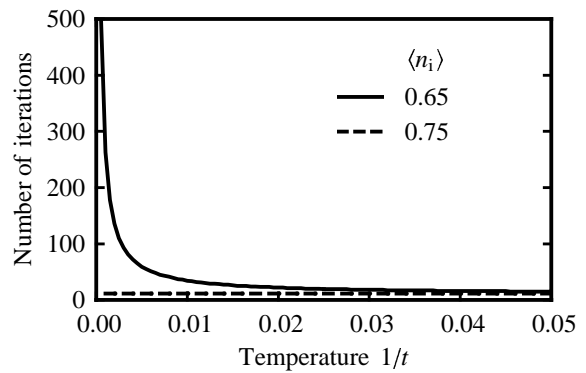} \vspace{1em}
\caption[Number of iterations required to reach a disordered phase
sink in the $d=2$\: $tJ$ model]{Number of iterations required to
reach a disordered phase sink in the $d=2$ isotropic $tJ$ model,
plotted as a function of temperature for two different values of
$\langle n_i \rangle$.  The value $<n_i> = 0.65$ is in the $\tau$
range.}\label{anis_fig:6}
\end{figure}

In contrast to $d=3$ even with the weakest coupling between planes,
in $d=2$ the $\tau$ phase sink is not a true sink fixed point of the
recursion relations, but it is a "quasisink" in the sense that
renormalization-group flows come close, stay in its vicinity for
many iterations, before crossing over along the disorder line to one
of the disordered sinks.  We thus find a zero-temperature $\tau$
critical point (which emerges from zero temperature with the
slightest inclusion of interplanar coupling, leaving behind a true
sink). The quasisink behavior is particularly true for trajectories
initiating at low temperatures, where the quasisink that is reached
is, numerically, essentially indistinguishable from a real one.
Since regions of the phase diagram that are approximately basins of
attraction of the quasisink are characterized by a sharp rise in the
number of iterations required to eventually reach the disordered
sinks, we can extract useful information by counting these
iterations.

We choose a numerical cutoff for when the interaction constants in
the rescaled Hamiltonian have come sufficiently close to their
limiting values at any of the high-temperature disordered fixed
points (the dilute disordered sink, the dense disordered sink, or
the null fixed point in-between). We then count the number of
iterations required to meet this cutoff condition for a given
initial Hamiltonian. Fig.~\ref{anis_fig:5} shows the results as
contour diagrams, plotted in terms of temperature vs. chemical
potential and temperature vs. electron density.  There are two clear
regions in Fig.~\ref{anis_fig:5}(a) where the number of iterations
blows up at low temperatures.  The region for $\mu/J$ approximately
between -0.5 and 1.6 flows to the $\tau$ phase quasisink. When
expressed in terms of electron density in Fig.~\ref{anis_fig:5}(b),
this region is centered around a narrow range of densities near
$\langle n_i \rangle = 0.65$, which is where the $\tau$ phase
actually emerges for finite $t_z/t_{xy}$. The low-temperature region
for $\mu/J \gtrsim 1.6$ flows to an antiferromagnetic quasisink, but
does not appear in the electron density contour diagram because the
entire region is mapped to $\langle n_i \rangle$ infinitesimally
close to 1.  This is similar to what we see in the anisotropic model
for low $t_z/t_{xy}$, where the antiferromagnetic region is stable
to only very small hole doping away from $\langle n_i \rangle = 1$,
but gradually spreads to larger doping values as the interplane
coupling is increased.  Fig.~\ref{anis_fig:6} shows the
zero-temperature $\tau$ fixed point behavior in another way, by
plotting the number of renormalization-group iterations as a
function of temperature, for two different $\langle n_i \rangle$.
For $\langle n_i \rangle = 0.65$, in the $\tau$ phase range, the
number of iterations diverges as temperature is decreased.  In
contrast, for $\langle n_i \rangle = 0.75$, not in the $\tau$ phase
range, the number is nearly constant at all temperatures.  In
summary, we see that the $d=2$ results are compatible with the small
$t_z/t_{xy}$ limit of the anisotropic model. A weak interplane
coupling stabilizes both the $\tau$ and antiferromagnetic phases,
yielding finite transition temperatures.

\begin{figure}
\centering

\includegraphics*[scale=1]{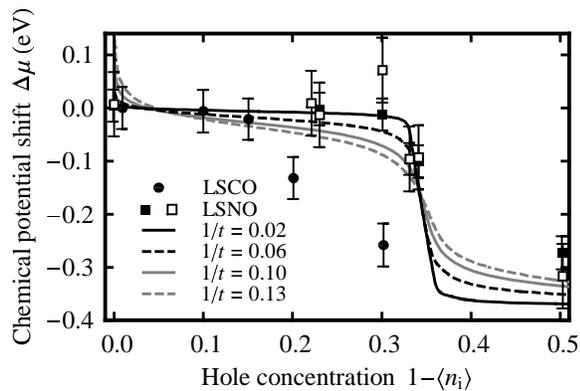} \vspace{1em}

\caption[The calculated chemical potential shift $\Delta \mu $
plotted as a function of hole concentration $1-\langle n_i \rangle$
for the $d=2$ \: $tJ$ model]{The calculated chemical potential
shift $\Delta \mu $ is plotted as a function of hole concentration
$1-\langle n_i \rangle$ for the isotropic $d=2$ $tJ$ model, at four
different temperatures.  For comparison with experimental results, the
energy scale $\tilde{t} = 0.1$ eV is chosen.  With this scale, the
temperatures $1/t = 0.02$, $0.06$, $0.10$ and $0.13$ correspond to 23,
70, 116, and 151 K respectively.  Experimental values for $\Delta \mu$
determined from x-ray photoemission spectra at $\sim 80$ K are shown
for the cuprate La$_{2-x}$Sr$_x$CuO$_{4}$ (LSCO, filled
circles)~\cite{AIno} and the nickelate La$_{2-x}$Sr$_x$NiO$_{4}$
(LSNO, filled squares)~\cite{ASatake}.  For LSNO we also show another
experimental estimate based on ultraviolet photoemission spectra (open
squares), taken at 150 K, except for the datapoint at zero hole
concentration, which was taken at 230 K~\cite{ASatake}.}\label{anis_fig:7}

\end{figure}

We mentioned earlier that the lamellar structure of A and D phases
which appears in the anisotropic $tJ$ phase diagram for hole dopings
up to the $\tau$ phase might be an indicator of incommensurate
ordering.  One possible form this incommensurate ordering could
take is the appearance of stripes, the segregation of the holes into
D-like stripes where the hole kinetic energy is minimized,
alternating with A-like stripes of antiferromagnetic order.
Depending on the arrangement of such stripes with respect to the
underlying lattice, the system could flow under repeated
renormalization-group transformations either to the
antiferromagnetic or dense disordered sink.  Since the arrangement
of the stripes will vary as we change the temperature or density in
the system, this could lead to a lamellar structure of A and D
phases in the resulting phase diagram.  Though we cannot probe the
existence of such stripes directly in our approach, an observable
consequence of stripe formation would be the suppression of the
chemical potential shift when additional holes are added to the
system, since we effectively have a phase separation on a
microscopic scale into hole-rich and hole-poor regions.  Indeed,
inquiries into stripe formation in experimental systems doped away
from half-filling often look for this tell-tale pinning of the
chemical potential.  For example, in the cuprate superconductor
La$_{2-x}$Sr$_x$CuO$_{4}$ (LSCO), photoemission measurements of core
levels have shown that the chemical potential shifts by a small
amount ($< 0.2$ eV/hole) in the underdoped region, $\delta \equiv
1-\langle n_i \rangle \lesssim 0.15$, compared to a large shift
($\sim 1.5$ eV/hole) in the overdoped region, $\delta \gtrsim 0.15$,
an observation which has been interpreted as a possible signature of
stripes~\cite{AIno}.  In non-superconducting systems where the
existence of stripes is clearly established, like the nickelate
La$_{2-x}$Sr$_x$NiO$_{4}$ (LSNO), we see a qualitatively similar
behavior, with the chemical potential shifting significantly only
for high-doping ($\delta \gtrsim 0.33$ for LSNO)~\cite{ASatake}. For
the $tJ$ model, we take the chemical potential shift as $\Delta \mu
= \tilde\mu - \tilde\mu_0$, where $\tilde\mu_0$ is the chemical
potential below which $\langle n_i \rangle$ begins to the decrease
noticeably from 1 in the low temperature limit.
Fig.~\ref{anis_fig:7} shows our calculated $\Delta\mu$ vs. hole
concentration for the $d=2$ $tJ$ model at four different
temperatures. In order to compare with the experimental data for
LSCO and LSNO, we choose an energy scale $\tilde t = 0.1$ eV.  For
the low-doping region, where interplane coupling generates a
lamellar structure of A and D phases, the slope of the $\Delta \mu$
curve remains small. On the other hand, for high-doping, in the
range of densities corresponding to the $\tau$ phase, $\Delta\mu$
turns steeply downward.  The similarities between this behavior and
the experimental data supports the idea of stripe formation in the
low-doping region.

\begin{acknowledgments}
This research was supported by the U.S. Department of Energy under
Grant No. DE-FG02-92ER-45473, by the Scientific and Technical
Research Council of Turkey (T\"UBITAK), and by the Academy of
Sciences of Turkey.  MH gratefully acknowledges the hospitality of
the Feza G\"ursey Research Institute and of the Physics Department
of Istanbul Technical University.
\end{acknowledgments}

\end{document}